\documentclass[prl,reprint,amsmath,amssymb,aps]{revtex4-1}
\pdfoutput=1
\usepackage{fullpage}
\usepackage{graphicx}
\usepackage{graphicx}
\usepackage{dcolumn}
\usepackage{bm}
\newcommand{\be}{\begin{equation}}
\newcommand{\ee}{\end{equation}}
\newcommand{\bea}{\begin{eqnarray}}
\newcommand{\eea}{\end{eqnarray}}

\newcommand{\bnabla}{\mbox{\boldmath $\nabla$}}
\renewcommand{\epsilon}{\varepsilon}

\begin{document}
\title{
The Length of Excitable Knots
}
\author{Fabian Maucher$^{\dagger\star}$ and Paul Sutcliffe$^\star$\\ \ }
\affiliation{
  $^\dagger$Joint Quantum Centre (JQC) Durham-Newcastle, Department of Physics,
Durham University, Durham DH1 3LE, United Kingdom.\\
  $^\star$Department of Mathematical Sciences,
Durham University, Durham DH1 3LE, United Kingdom.\\ 
Email: fabian.maucher@durham.ac.uk, \ p.m.sutcliffe@durham.ac.uk}
\date{June 2017}

\begin{abstract}
  The FitzHugh-Nagumo equation provides a simple mathematical model
  of cardiac tissue as an excitable medium hosting spiral wave vortices.
  Here we present extensive numerical simulations studying long-term dynamics of knotted vortex string solutions for all torus knots up to crossing number 11.
  We demonstrate that FitzHugh-Nagumo evolution preserves the knot topology
  for all the examples presented,
  thereby providing a novel field theory approach to the study of knots.
  Furthermore, the evolution
  yields a well-defined minimal length for each knot that is comparable
  to the ropelength of ideal knots. We highlight
  the role of the medium boundary in stabilizing the length of the knot and
  discuss the implications beyond torus knots. By applying Moffatt's test we are able to show that there is not a unique attractor within a given knot topology.
\end{abstract}
\maketitle

There are a range of chemical, physical and biological excitable media
that support spiral wave vortices. Examples include the
Belousov-Zhabotinsky redox reaction, the chemotaxis of slime mould and action potentials in cardiac tissue \cite{Winfree}.
Spiral waves in the latter system are
of particular importance as they are believed to play a vital role in certain
cardiac arrythmias \cite{Wit}. The simplest mathematical model of cardiac tissue as
an excitable medium is the
FitzHugh-Nagumo equation \cite{FH,Nag,Kog}, which is a nonlinear partial differential equation
of reaction-diffusion type for the electric potential together with
a slow recovery variable.
In a three-dimensional medium spiral wave
vortices become extended vortex strings (sometimes known as scroll waves)
which can either end on the boundary of the medium or form closed loops.
Mathematically, such a closed loop is a knot, which includes the trivial
case of the unknot as a vortex ring.
\begin{figure}
\includegraphics[width=0.97\columnwidth]{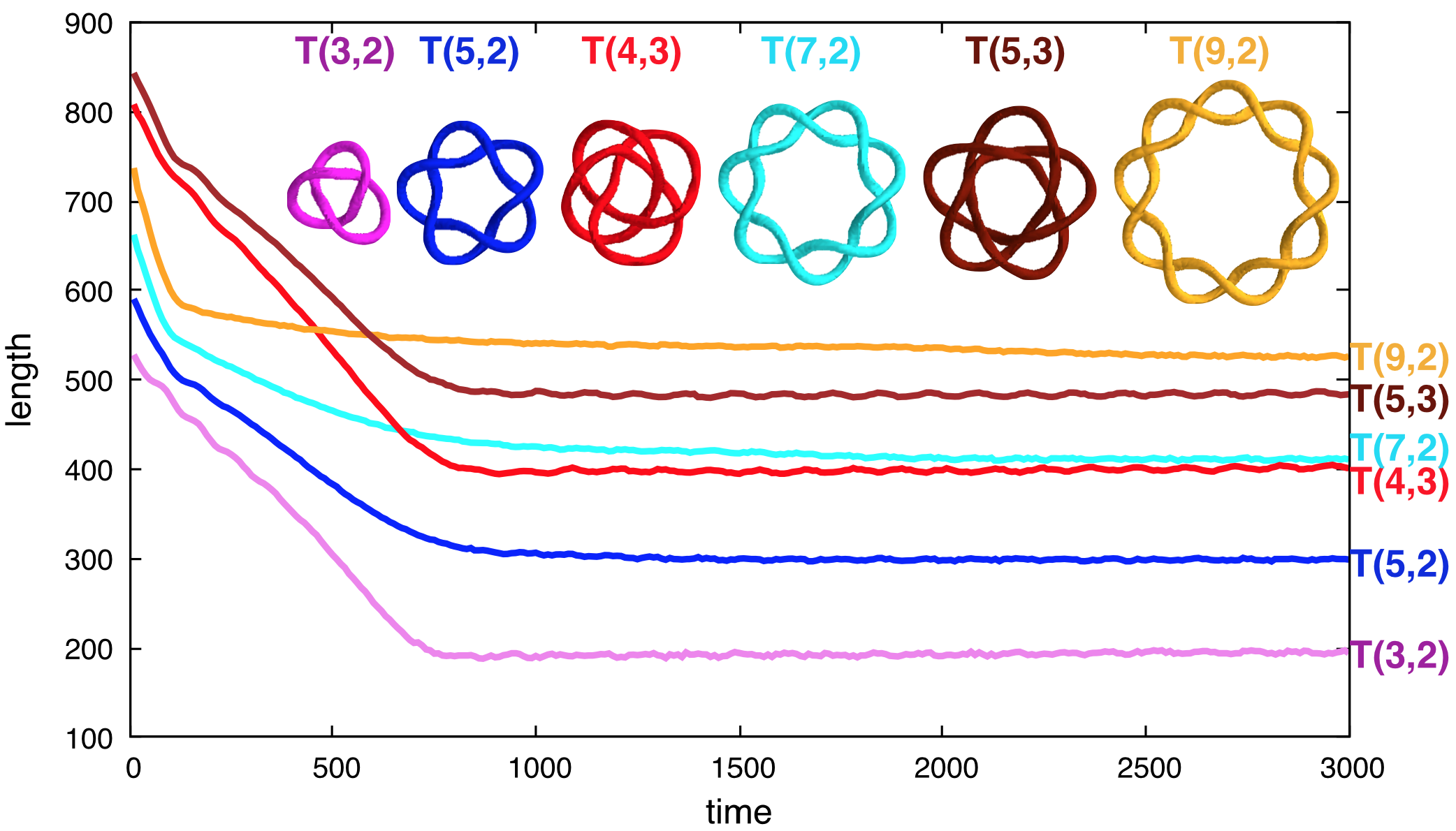}
 \includegraphics[width=0.97\columnwidth]{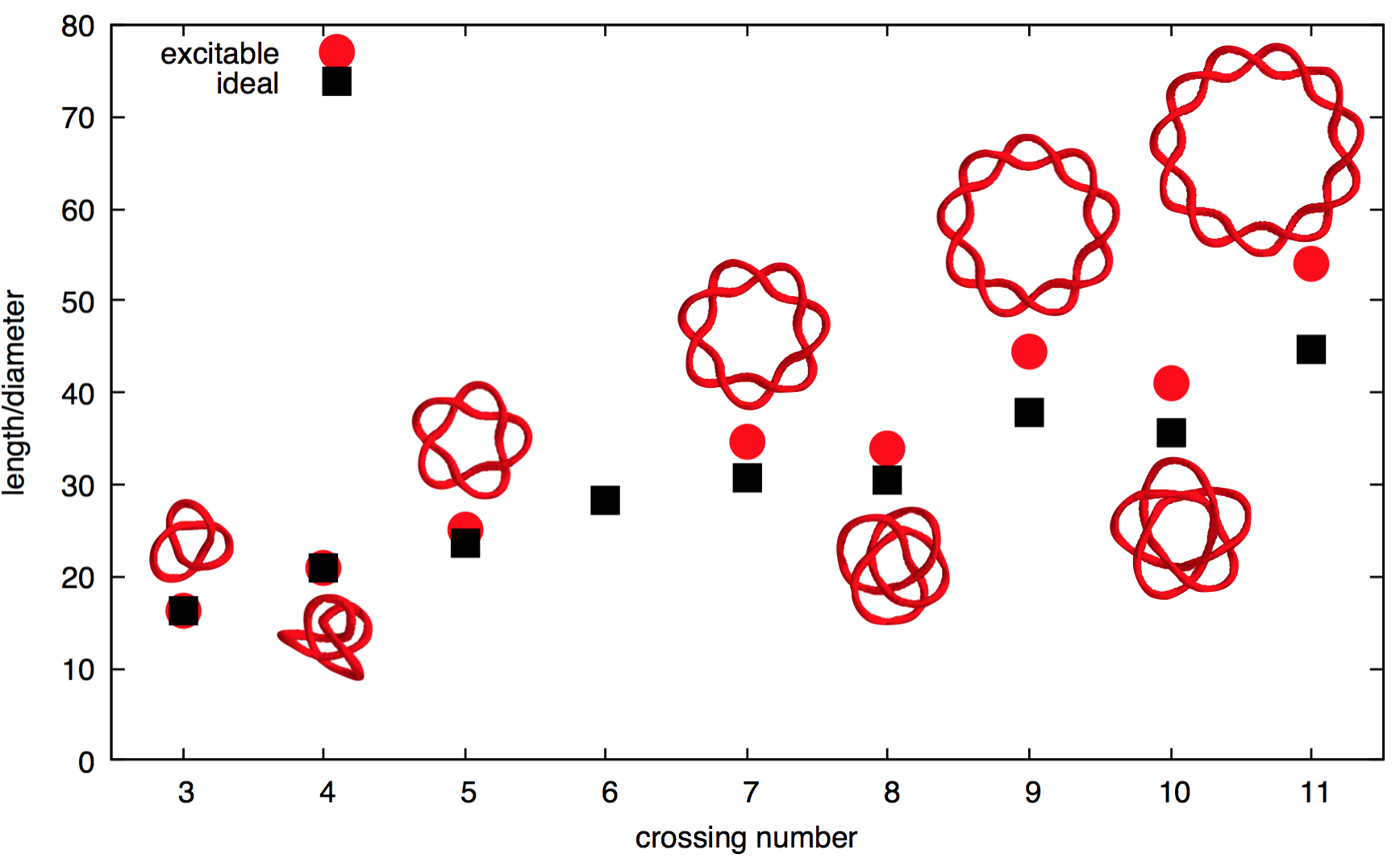}
 \caption{The upper panel shows the evolution of the length of some torus knots. The lower panel displayes the length/diameter (circles) as a function of crossing number and the ropelength of ideal knots (squares).}   
\label{fig1}
\end{figure}

Over thirty years ago it was conjectured \cite{StWi} that non-trivial knots 
in a FitzHugh-Nagumo medium might preserve their topology
and be remarkably immune to the
reconnection events that untie knotted vortex strings in a generic way
in most systems \cite{KKI}. To date, the only evidence \cite{SW} over long time
scales for this conjecture has been restricted to the simplest
non-trivial knot, the trefoil knot, and some very recent results on the
untangling of unknots \cite{MS}.
Here we provide significant new evidence for this conjecture
 by presenting solutions for all torus knots up to
crossing number 11. We find that the knot topology is preserved and 
the evolution yields a stable minimal length for each knot.
The combination of positive string tension and short-range
repulsion between vortex cores leads to the
naive expectation that the length of these excitable knots is
related to the ropelength of ideal knots \cite{Kat}, a concept introduced to explain the properties of knotted DNA.
We investigate this possible connection to ideal knots and
find that these two lengths follow similar trends. 
Differences in conformation and the crucial role of the medium boundary
in stabilizing the length of the knot are highlighted.
This latter phenomenon is relevant for extending the results beyond torus knots. Finally, we also demonstrate that there
is not a unique attractor within a given knot topology, by applying Moffatt's test \cite{Mo} to the trefoil knot.

The FitzHugh-Nagumo medium is described by the nonlinear
reaction-diffusion partial differential equations 
\be
\frac{\partial u}{\partial t}=\frac{1}{\epsilon}(u-\frac{1}{3}u^3-v)+\nabla^2 u,
\quad
\frac{\partial v}{\partial t}=\epsilon(u+\beta-\gamma v),
\label{FHN}
\ee
where $u({\bf r},t)$, represents the electric potential and
$v({\bf r},t)$ is the recovery variable, both being real-valued physical fields defined throughout the three-dimensional medium with spatial coordinate ${\bf r}$ and time $t$.
The remaining variables are constant parameters that we fix to be
$\epsilon=0.3, \ \beta=0.7, \ \gamma=0.5$ from now on, to avoid complications
due to spiral wave meander \cite{Win2}.
In two-dimensional space this system, with the parameter values given above,
has rotating spiral wave vortex solutions \cite{Winfree},
with a period ${\cal T}=11.2$ and $u$ and $v$ wavefronts in the form of an involute spiral with a wavelength $\lambda=21.3.$ Characteristic time and length scales
for the system are provided by the parameters ${\cal T}$ and $\lambda$, respectively.
The centre of the vortex
is the point at which $|{\bnabla} u\times \bnabla v|$ is maximal,
and this quantity is 
localized in the vortex core \cite{Win3}.

To provide initial conditions for a vortex string of an
arbitrary knot, given by any
non-intersecting closed curve $K$, we apply the method introduced in \cite{MS} and adapted from \cite{Sa}. This involves computing the initial fields $u({\bf r},0)$ and $v({\bf r},0)$ from a scalar potential for a vector field defined
by a Biot-Savart integral along the curve $K$. For most of our investigations we shall restrict to the case where $K$ is a torus knot.
By definition, torus knots can be restricted to lie on the surface of a torus
and are classified by a pair of coprime integers $p>q>1$, with the associated
torus knot denoted by $T(p,q).$ A suitable explicit parametrization for the curve $K$
can be taken to be
\be
   {\bf r}=\begin{pmatrix}(R_1+R_2\cos(p\phi))\cos(q\phi)\\
     (R_1+R_2\sin(p\phi))\sin(q\phi)\\
   -R_2\sin(p\phi)
   \end{pmatrix}
   \label{par}
\ee
where $\phi\in[0,2\pi)$ and $R_1>R_2$ are the major and minor radii of the torus
  $(R_1-\sqrt{x^2+y^2})^2+z^2=R_2^2.$ To ensure that the initial conditions
  generate a vortex string with segments no closer than about half a spiral wavelength (to avoid initial reconnection), we generally found that $R_1=20$ and $R_2=10$ were sufficient, although $R_1=40$ was used for some larger values of $p$.
  The above explicit parametrization shows that $p$ and $q$ are the number of times that the knot winds around the poloidal and toroidal directions of the torus respectively. The knots $T(p,q)$ and $T(q,p)$ are topologically equivalent, as the former may be smoothly deformed into the latter, hence the restriction $p>q$ in the topological classification of torus knots.
  The (minimal) crossing number of the $T(p,q)$ torus knot is $p(q-1)$, and
  for all crossing numbers from 3 to 11 there is a unique torus knot except
  for crossing numbers 4 and 6, where torus knots do not exist.

  The FitzHugh-Nagumo equations (\ref{FHN}) are solved in a spatial region
  that is a cuboid of height 64 and length and width equal to 128 (increased to 192 for the longest knots), so that the cuboid spans
  at least a few spiral wavelengths in each direction.
  No-flux (Neumann) boundary conditions are imposed at all boundaries of the medium and standard numerical methods are employed: fourth-order Runge-Kutta time evolution with a timestep $0.1$ and a 27 point stencil finite difference approximation for the Laplacian with a lattice spacing $0.5.$
  The vortex string is visualized by plotting an isosurface where $|{\bnabla} u\times \bnabla v|$ takes the value 0.1. This surface is a hollow tube and the vortex string is defined to be the curve that is the centreline of this tube.
  To calculate the length of a knot we first identify the lattice point where  $|{\bnabla} u\times \bnabla v|$ is maximal and then compute a sequence of neighbouring lattice points by employing a path following algorithm that selects the neighbour with the greatest value of $|{\bnabla} u\times \bnabla v|$ and terminates once it returns to the original starting point. A discrete Fourier transform is then performed on this sequence to obtain a finite Fourier series representation of a smooth curve, with a length that is easily computed from the Fourier coefficients. We used a Fourier series representation with 50 terms and
verified that the computed length is insensitive to any reasonable
variation in the number of terms around this value. 

 The upper panel in Fig.~\ref{fig1} displays the evolution of the length of the knot as a function of time using the initial conditions described above (with appropriate values of $p$ and $q$) for all torus knots up to crossing number 10. In each case, we find that the knot topology is preserved (there are no reconnections) and the length has stabilized to an approximately constant value after a time of $3000$, which is a few hundred spiral periods. The knots at this final time are also displayed, to scale and in order of increasing length, by plotting
 the isosurface given by $|{\bnabla} u\times \bnabla v|=0.1.$
 The knot $T(11,2)$, with crossing number 11, has also been computed, but for clarity this is not shown in the upper panel, as the evolution takes slightly longer to reach an equilibrium length. The evolution of this example is available as the movie {\em t112.mpg} in the supplementary material, together with the
 example $T(5,3)$ as the movie  {\em t53.mpg}.
 The supplementary material makes clear the dynamical nature of these knots and highlights the small oscillation in knot length around an equilibrium value as the vortex string slowly rotates and breaths with a period much greater than the spiral period. These results reveal that the amplitude of the oscillation in the length decreases as $p$ increases but increases as $q$ increases.

 It has been shown that FitzHugh-Nagumo flow is able to untangle unknots with a complex initial conformation \cite{MS} but the physical mechanism that underpins this process remains elusive. Although the untangling dynamics is highly nontrivial, a phenomenological understanding, applicable in the regime of slight curvature and twist, can be obtained by combining positive filament tension \cite{Ke,BHZ} with a short-range repulsion between vortex strings.
  If this simplified description has merit then it leads to the expectation that the length of an excitable knot should bear a resemblance to the ropelength of an ideal knot.
  Consider a perfectly flexible tube with a cross-section that is a rigid disk of unit diameter. The centreline of the tube is an ideal knot if the tube has minimal length within a given knot topology and this length is the ropelength of the knot \cite{Kat}. Extensive data on the computed ropelengths of knots can be found in \cite{Ash}. As ropelength is defined as the ratio of length to diameter then to fix a normalization we need to define the diameter of our excitable knots. Here we choose to fix units by matching to the ropelength of the simplest knot, the trefoil knot $T(3,2)$, also denoted $3_1$ in Alexander-Briggs notation where a knot is labelled by its crossing number with a subscript that denotes its position in the Rolfsen knot table. This yields a diameter of 11.8, which is around half a spiral wavelength.

\begin{figure}
 \includegraphics[width=\columnwidth]{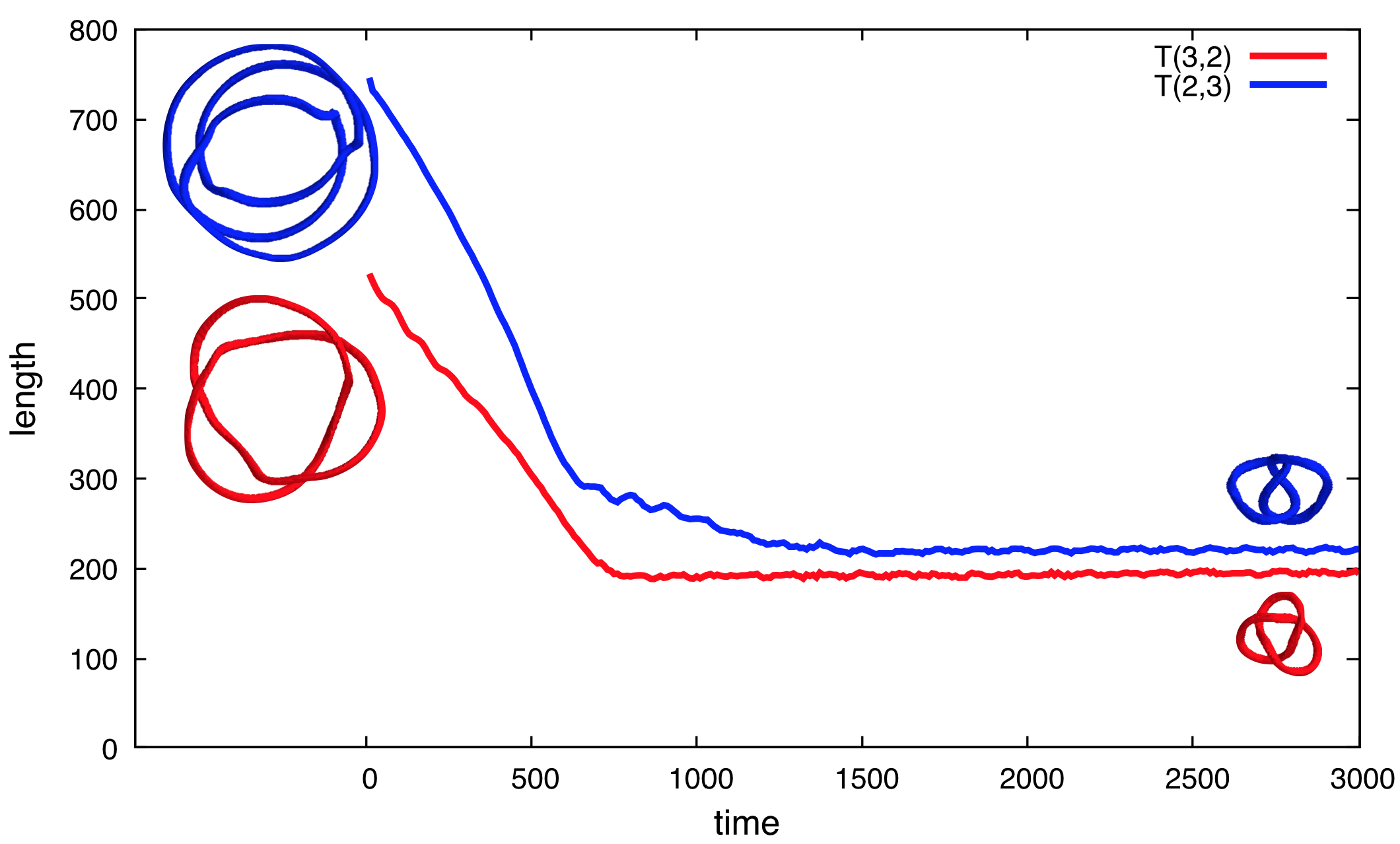}
 \caption{Lengths and snapshots (to scale) of the evolution of the
   topologically equivalent initial knots $T(3,2)$ (lower red curve) and $T(2,3)$ (upper red curve).
   This shows that FitzHugh-Nagumo flow fails 
Moffatt's test, as $T(2,3)$ fails to transmute into $T(3,2)$.}
\label{fig-mof}
\end{figure}
  
In the lower panel in Fig.~\ref{fig1} the circles show the ratio of length to diameter for all excitable torus knots with crossing number up to 11 and for comparison the squares show ropelength of the same ideal knots. This plot reveals that the ropelength gives a reasonable first estimate of the length of an excitable knot, despite its simplicity. Some features of ideal knots are reproduced by excitable knots, for example, the surprising fact that the ideal knot $8_{19}$ (the torus knot $T(4,3)$) has a ropelength that is less than that of any knot with crossing number seven is replicated by excitable knots. 
Note that we have included the figure-eight knot $4_1$, the only knot with crossing number four, in the data in Fig.~\ref{fig1}, even though this is not a torus knot. We shall discuss this example in more detail later.

Although the lengths of excitable knots are in reasonable agreement with the predictions from ideal knots, their conformations are quite different.
This naturally leads to the question of whether
  FitzHugh-Nagumo flow has a unique attractor within a given knot topology.
  In the context of numerical algorithms to find ideal knots,
  Moffatt \cite{Mo} proposed the astute test to start with the $T(2,3)$ conformation of the trefoil knot to see if it transmutes into the $T(3,2)$ form.
  In Fig.~\ref{fig-mof} we present the results of applying Moffatt's test to
  FitzHugh-Nagumo flow. The initial configuration and the resulting
  solution at time $3000$ are shown to scale for both cases, together with plots
  of the evolution of the length as a function of time. These results show that in each case the length stabilizes around a constant value and an equilibrium conformation is attained, but as the two are different the flow fails Moffatt's test and we have demonstrated that there is not a unique attractor within a given knot topology. Note that the conformation of the slightly longer form of $3_1$
  is quite symmetric and is very different from the initial condition.

\begin{figure}
 \includegraphics[width=\columnwidth]{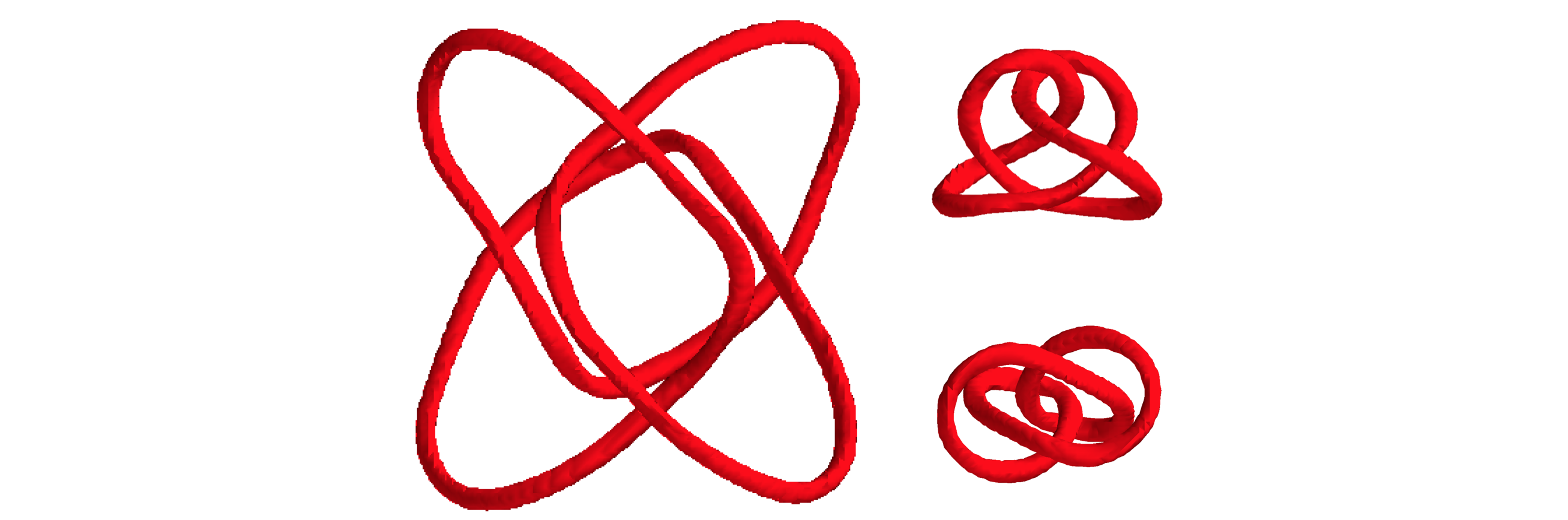}
 \caption{On the left is the initial condition for the figure-eight knot $4_1$, obtained by scaling the ideal knot, and on the right are two views of the final configuration.
}
\label{fig-41}
\end{figure}

  As there is not a unique attractor,
  we cannot rule out the possibility that the ideal knot is also an attractor, which could have been obtained with more favourable initial conditions.
An obvious choice for an alternative initial condition is to take an ideal knot, suitably scaled so that initially all segments are sufficiently far apart.
The initial conditions for ideal knots were obtained using the data available at \cite{webpage} for the associated curves $K$.
In Fig.~\ref{fig-41} the left image is the 
initial condition for the figure-eight knot $4_1$ in ideal form and the
images on the right are two views of the final configuration
after FitzHugh-Nagumo flow for a time of $3000$, which 
yields a rather symmetric but non-ideal conformation.
This same solution is also obtained from other initial conditions,
including the more standard figure-eight form with the minimal number of four crossings, demonstrating that the final state is robust with respect to the initial condition.

\begin{figure}
 \includegraphics[width=\columnwidth]{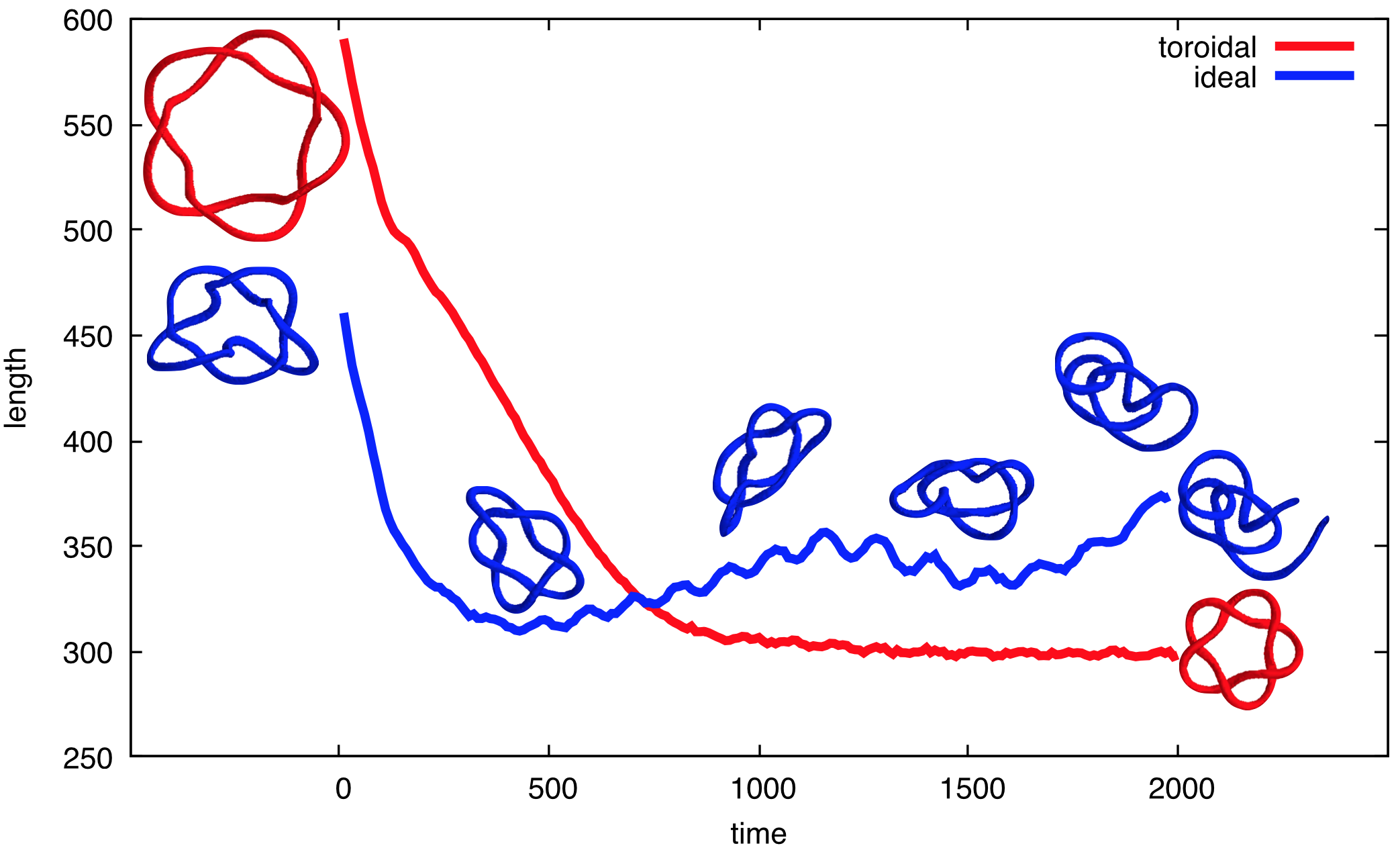}
 \caption{Lengths and snapshots (to scale) of the evolution of the knot $T(5,2)$
with an initial toroidal conformation (red curve) and a scaled
 ideal conformation (blue curve). 
 Although both knots initially shrink, only the toroidal conformation
 nestles against the medium boundary (lower snapshot at time $2000$)
 and attains an equilibrium length.
 The initially ideal conformation displays an instability and
 even though the topology is preserved this knot breaks at the medium boundary
 (final upper snapshot at time $1980$) before an equilibrium length could be realized.}
\label{fig-51}
\end{figure}

The example of the ideal knot $5_1$ (the torus knot $T(5,2)$) is presented in Fig.~\ref{fig-51}, where the blue curve displays the evolution of length with time and insets show the initial condition (on the left) and snapshots to scale at
 increasing times. For comparison, the length of the toroidal form of the same knot is also presented (red curve) together with initial and final snapshots. We find that the evolution of the intially ideal form again preserves topology (there are no reconnections) and the length initially decreases to a reasonably low value. However, the length does not stabilize and begins to increase, with the knot taking an ever more complicated conformation until it eventually breaks in a collision with the boundary of the medium at time $1980$. Similar three-dimensional instabilities have been investigated in detail \cite{HH} for the case of a single initially straight vortex string in a similar excitable medium, where it has been shown that twist can induce instability.

 The reason that the toroidal forms of the torus knots have a stabilized length is that they drift along the axis of the torus until they reach the boundary of the medium, where their compact form allows them to nestle snugly against the no-flux boundary. In the case of a circular vortex ring in an excitable medium it has been observed both numerically \cite{NW} and experimentally \cite{ATE} that
 close proximity to the medium boundary can suppress instabilities. Furthermore, the same small amplitude breathing mode that we observe in the long-time evolution of the knot length is also found for a boundary stabilized ring \cite{ATE}, providing a smoking gun signal of the crucial role of the medium boundary.
 Moreover, we have performed simulations of initially toroidal forms in cuboids with a much greater height, and found that an instability in the length can emerge over a timescale of the order of a few hundered spiral periods, if the knot is not able to drift all the way to the medium boundary over this timescale.

 In previous simulations \cite{SW} of the trefoil knot in the FitzHugh-Nagumo medium, periodic boundary conditions were employed in the direction of the initial drift with no-flux boundary conditions in the other directions. In that case it was found that an instability of the length develops over a timescale of a few hundred spiral periods but over much longer timescale, of the order of thousands of spiral periods, the length recovers to the same value found in the present study. We are now in a position to explain this finding, in that the increasing length is accompanied by a dramatic (but topology preserving) change in conformation that changes the drift direction so that the knot eventually makes it to one of the no-flux boundaries, where the instability is suppressed and the knot recovers its minimal length.
 So far, we have found that for knots with crossing number five or more, the twist required for such crossing numbers means that an instability develops unless the initial condition is sufficiently close to the compact toroidal form and close to the no-flux medium boundary. It is not known whether knots with these initial instabilities might eventually recover their length, like the earlier trefoil example, since all simulations conducted so far end in the knot breaking at the medium boundary. It is difficult to resolve this issue because of the 
 Catch-22 situation that a larger simulation region is required to avoid the knot breaking at the boundary but a boundary that is further from the knot cannot suppress the instability. These difficulties have so far prevented the computation of any stabilzed forms of knots with six crossings, where there are no torus knots.

 In summary, we have computed a range of knots in the FitzHugh-Nagumo excitable medium and compared their properties to ideal knots. We find that knot topology is preserved by the flow, thereby providing a novel field theory approach to knots that complements both traditional material models of knots and more abstract mathematical concepts. As knots play an increasingly important role in a variety of contexts, for example in the study of DNA \cite{WC},
 a new approach to knots, as solutions of
 nonlinear partial differential equations, may find applications in a variety of areas.\\

 \ 
 
  {\bf \small \qquad \qquad ACKNOWLEDGEMENTS}
  
\noindent  We thank Gareth Alexander for useful discussions.
This work is funded by the
Leverhulme Trust Research Programme Grant RP2013-K-009, SPOCK: Scientific Properties Of Complex Knots, and the STFC grant ST/J000426/1.


\begin{thebibliography}{99}

\bibitem{Winfree} A.T. Winfree, \textit{The Geometry of Biological Time},
  Springer-Verlag, 2001.

\bibitem{Wit}  
   F.X. Witkowski {\em et al},
   \textit{Nature} {\bf 392}, 78 (1998).

 \bibitem{FH}
   R. FitzHugh,
   \textit{Biophys. J.} {\bf 1}, 445 (1961).

 \bibitem{Nag}  
   J.S. Nagumo, S. Arimoto and S. Yoshizawa,
   \textit{Proc. IRE.} {\bf 50}, 2061 (1962).

 \bibitem{Kog} B.Y. Kogan, \textit{Introduction to computational cardiology},   
Springer, 2010.
   
\bibitem{StWi} A.T. Winfree and S.H. Strogatz,
  \textit{Nature} {\bf 311}, 611 (1984).
  
  \bibitem{KKI} D. Kleckner, L.H. Kauffman and W.T.M. Irvine,
    \textit{Nat. Phys.} {\bf 12}, 650 (2016).

\bibitem{SW} P.M. Sutcliffe and A.T. Winfree,
  \textit{Phys. Rev. E} {\bf 68}, 016218 (2003).    

\bibitem{MS} F. Maucher and P.M. Sutcliffe,
  \textit{Phys. Rev. Lett.} {\bf 116}, 178101 (2016).

\bibitem{Kat}
 V. Katritch {\em et al},
  \textit{Nature} {\bf 384}, 142 (1996).
  
\bibitem{Mo} H.K. Moffatt,
  \textit{Nature} {\bf 384}, 114 (1996).

 \bibitem{Win2}
 A.T. Winfree,
 \textit{Chaos} {\bf 1}, 303 (1991).

 \bibitem{Win3}
 A.T. Winfree,
 \textit{Physica D} {\bf 84}, 126 (1995). 
  
\bibitem{Sa} H. Salman,
  \textit{Phys. Rev. Lett.} {\bf 111}, 165301 (2013).

\bibitem{Ke} J.P. Keener,
  \textit{Physica D} {\bf 31}, 269 (1988).

\bibitem{BHZ} V.N. Biktashev, A.V. Holden and H. Zang,
  \textit{Philos. Trans. R. Soc. London} {\bf A347}, 611 (1994).
    
\bibitem{Ash}
  T. Ashton, J. Cantarella, M. Piatek and E.J. Rawdon,
  \textit{Experimental Math.} {\bf 20}, 57 (2011).

\bibitem{webpage}
 Knot Atlas: \verb|http://katlas.org/wiki/Ideal_knots|
  
\bibitem{HH}
 H. Henry and V. Hakim
 \textit{Phys. Rev. E.} {\bf 65}, 046235 (2002).

 \bibitem{NW}
  P.J. Nandapurkar and A.T. Winfree,
  \textit{Physica D}, {\bf 35}, 277 (1989).

\bibitem{ATE}
A. Azhand, J.F. Totz and H. Engel,
\textit{EPL} {\bf 108}, 10004 (2014).

\bibitem{WC} S.A. Wasserman and N.R. Cozzarelli, \textit{Science} {\bf 232}, 951 (1986).

  
\end{thebibliography}
\end{document}